\documentclass[12pt,letterpaper]{article}
\usepackage{jheppub}
\usepackage{verbatim}
\usepackage{slashed}

\usepackage{graphicx}
\usepackage{subfigure}
\input{epsf}
\usepackage{epsfig}
\usepackage{epstopdf}
\usepackage{amsthm}

\def\({\left(} \def\){\right)}
\def\[{\left[} \def\]{\right]}
\def\al{\alpha} \def\bt{\beta}
\def\del{{\partial}}

\newcommand{\non}{\nonumber \\}

\newcommand{\be}{\begin{equation}}
\newcommand{\ee}{\end{equation}}
\newcommand{\bea}{\begin{eqnarray}}
\newcommand{\eea}{\end{eqnarray}}
\newcommand{\ba}{\begin{eqnarray}}
\newcommand{\ea}{\end{eqnarray}}

\newcommand{\beq}{\begin{equation}}
\newcommand{\eeq}{\end{equation}}
\newcommand{\beqa}{\begin{eqnarray}}
\newcommand{\eeqa}{\end{eqnarray}}
\newcommand{\beqar}{\begin{eqnarray*}}
\newcommand{\eeqar}{\end{eqnarray*}}

\newcommand{\reef}[1]{(\ref{#1})}

\newcommand{\eg}{{\it e.g.,}\ }
\newcommand{\ie}{{\it i.e.,}\ }

\newcommand{\A}{\mathcal{A}}

\title{Entanglement entropy, planar surfaces, and spectral functions}

\author{Vladimir Rosenhaus}
\emailAdd{vladr@berkeley.edu}
\affiliation{Center for Theoretical Physics and Department of Physics,\\
 University of California, Berkeley, CA 94720, U.S.A. }

\author{and Michael Smolkin}
\emailAdd{smolkinm@berkeley.edu}

\abstract{
We consider the universal part of entanglement entropy across a plane in flat space for a QFT, giving a non-perturbative expression in terms of a spectral function. We study the change in entanglement entropy under a deformation by a relevant operator, providing a pertrubative expansion where the terms are correlation functions in the undeformed theory. The entanglement entropy for free massive fermions and scalars easily follows. 
Finally, we study entanglement entropy across a plane in a background geometry that is a deformation of flat space, finding new universal terms arising from mixing of geometry and couplings of the QFT.
}

\begin{document}
\maketitle
 
\section{Introduction}

Entanglement entropy has emerged as a topic of interest in a wide range of areas \cite{Bomb86, Sred93, Callan94, CalCardy04,RefMoore04,KitPres05,RT}. Within the context of quantum field theory, entanglement entropy is a UV divergent quantity, with most of the contributions to it depending on the choice of UV regulator, $\delta$. However, in even space-time dimensions there is a term which is an exception, scaling like $\log(\delta)$. The coefficient of this term is the universal part of entanglement entropy, and is expected to encode the characteristics of the QFT.

On general grounds, one expects entanglement entropy to depend on the metric of the spacetime, the shape of the entangling surface, and the characteristics, such as the couplings, of the QFT. Furthermore, entanglement entropy will contain terms which depend solely on geometry (those that persist for a CFT), terms which depend solely on couplings (those that persist in the absence of extrinsic and background curvatures), and terms which involve mixing between the two. For instance, the exclusively geometric dependence was found by Solodukhin in \cite{solo,Fur13} for a $4$-dimensional space, with the universal part of entanglement entropy for a CFT expressed as an integral over the entangling surface of a combination of the induced metric, intrinsic and extrinsic curvatures (see also \cite{Hung:2011xb} for a $6$-dimensional CFT). 
In this paper, our primary concern will be the dependence of entanglement entropy exclusively on the characteristics of the QFT. To isolate this contribution, we consider a planar entangling surface in flat space. 

The starting point for our analysis is the entanglement entropy flow equation \cite{RS2}. This equation encodes the dependence of entanglement entropy on the coupling $\lambda$ of some relevant operator $\mathcal{O}$ in the field theory, 
\be \label{eq:SIntro}
\frac{\partial S}{\partial \lambda} = - \int d^d x\, \langle  K_{\lambda}\, \mathcal{O}(x) \rangle~_{\lambda}~.
\ee
Here $K_{\lambda}$ is the modular Hamiltonian, defined through the reduced density matrix $K_{\lambda} = -\log \rho_{\lambda}$, and $\langle \cdots \rangle$ denotes a connected correlation function in the vacuum of the theory with coupling $\lambda$. In Sec.~\ref{sec:linear} we review the derivation of this equation, and give an independent derivation of it through a direct expansion of the modular Hamiltonian. We exploit the fact that for a planar entangling surface in flat space, the modular Hamiltonian is known for any QFT and is simply the analytic continuation of the Rindler Hamiltonian \cite{KabStra, Bis75, Bis76, Crisp07}. Evaluation of entanglement entropy is therefore reduced to a computation of correlation functions in flat space. 

In Sec.~\ref{sec:allorders} we perform a perturbative expansion of both sides of (\ref{eq:SIntro}) in order to express the entanglement entropy for a theory with coupling $\lambda$ entirely in terms of correlation functions of a theory with coupling $\lambda_0$. In Sec.~\ref{sec:nonlinear} we assume there is one relevant coupling and work directly with Eq.~\ref{eq:SIntro} to express the entanglement entropy for a general theory in terms of its spectral function, and explicitly evaluate the entanglement entropy for a free theory.

In Sec.~\ref{sec:defgeom} we initiate a study of the interplay of background geometry with couplings of the theory. We find the coefficients, for a general QFT, of the universal $\log$ terms in entanglement entropy that are linear in the curvature. We achieve this by computing the first order change in the entanglement entropy of a plane in a background metric that is a small perturbation of flat space.

\section{Flat entangling surface: perturbative analysis} \label{sec:linear}

Consider some subregion $V$ of a manifold $\mathcal{M}$. The reduced density matrix for this region is obtained by tracing out degrees of freedom associated with $\overline{V}\,$-$\,$the complement of $V$,
\be \label{eq:rho}
\rho = \text{Tr}_{\overline{V}} |0\rangle \langle 0| \equiv {e^{-K}\over \text{Tr} \, e^{-K}} ~,
\ee
where we have taken the global state to be the vacuum. The right hand side of (\ref{eq:rho}) serves the definition of the modular Hamiltonian $K$. The entanglement entropy is defined as the von Neumann entropy of the reduced density matrix, 
\be \label{eq:Sdef}
S = -\text{Tr}_V\,(\rho\log\rho)~.
\ee

Let us assume that for a theory with action $I_0$ containing couplings $\lambda_0$ and defined on a general manifold $\mathcal{M}$ we know the reduced density matrix, $\rho_0$, for some entangling surface $\Sigma$. In what follows we address how the entanglement entropy changes if we slightly perturb the QFT by changing the coupling, $\lambda_0 \rightarrow \lambda_0 + \delta \lambda$, as well as how the entanglement entropy changes if instead we slightly deform the geometry of the background. This section, as well as section~\ref{sec:nonlinear}, will be concerned with the former, while section~\ref{sec:defgeom} will deal with the latter. Either of these deformations will lead to a change in the density matrix, 
\be
\rho = \rho_0 + \delta \rho~.
\ee
The resulting change in the entanglement entropy is found through an expansion of (\ref{eq:Sdef}). The first order term in $\delta \rho$ gives the so-called first law of entanglement entropy \cite{Bhattacharya:2012mi,Blanco:2013joa,Wong:2013gua},
\be \label{eq:firstlaw0}
\delta S = \text{Tr}(K_0 \delta \rho) ~.
\ee
The first law (\ref{eq:firstlaw0}) can alternatively be expressed in terms of the change $\delta K$ in the modular Hamiltonian \cite{Blanco:2013joa}, 
\be \label{eq:firstlaw}
\delta S = -\langle 0|\delta K \, K_0|0\rangle_0~.
\ee

\begin{figure}[tbp] 
\centering
\subfigure[]{
	\includegraphics[width=2in]{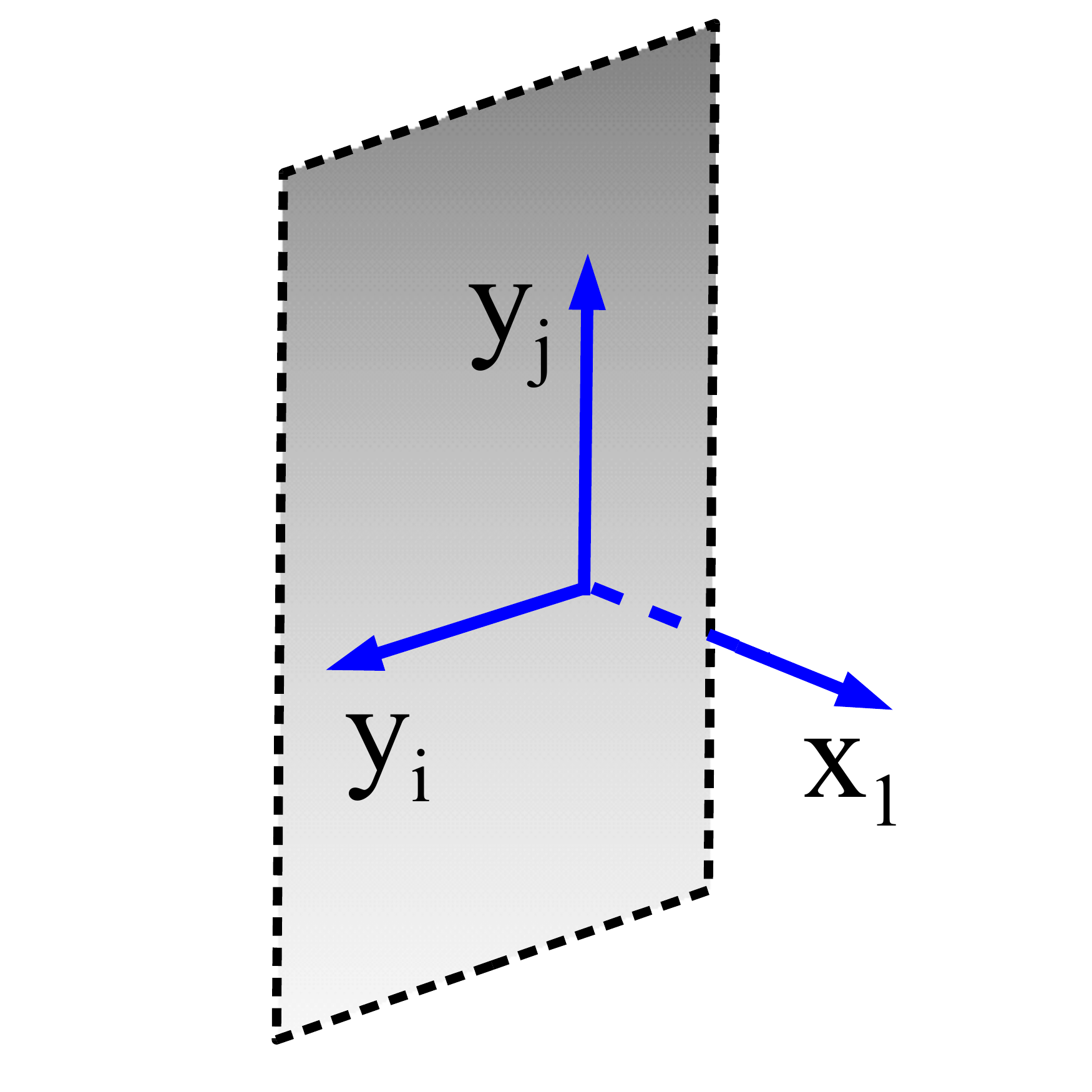}
	}
		\hspace{.2in}
		\subfigure[]{
	\includegraphics[width=2in]{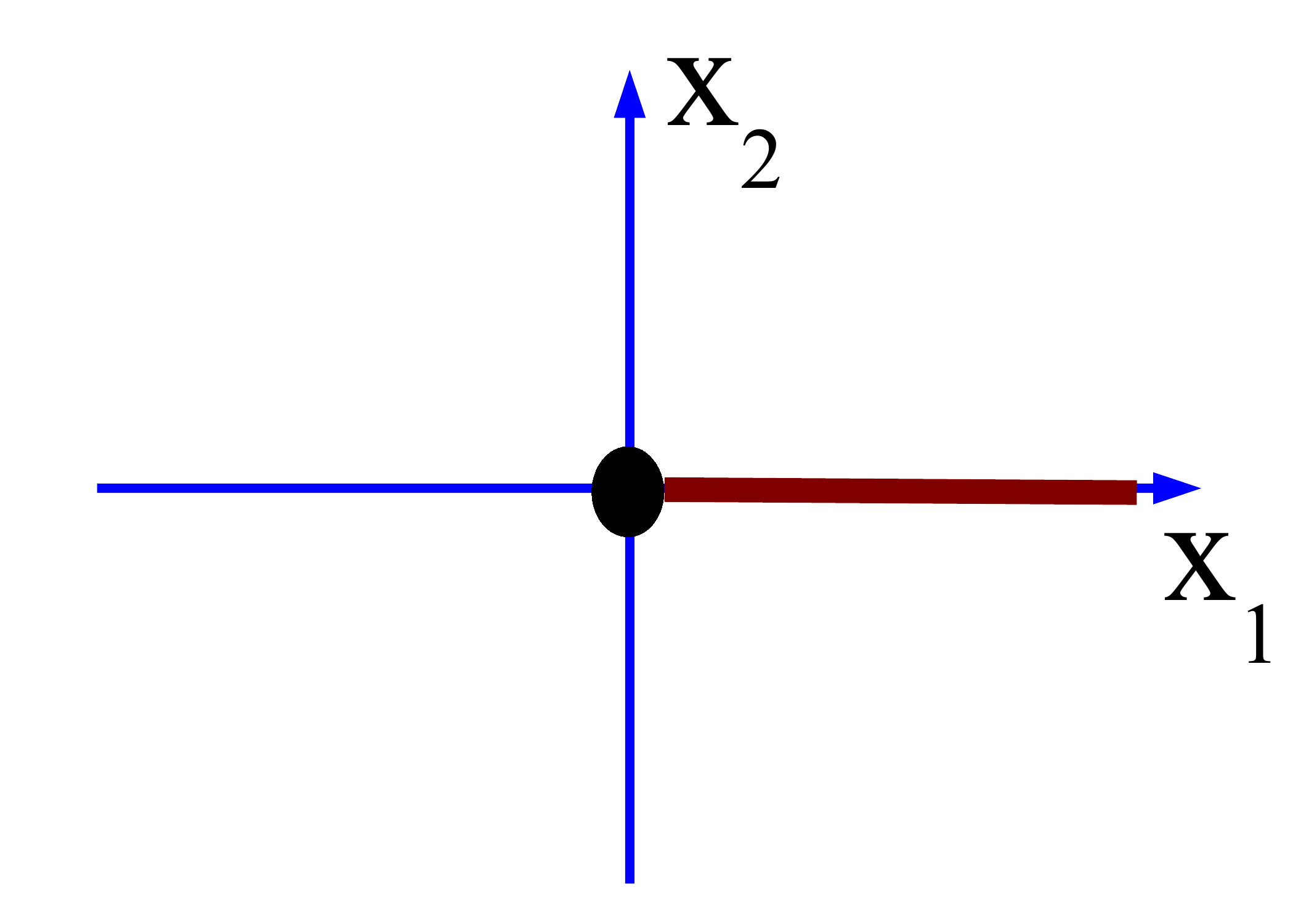}
	}
\caption{(a) An entangling surface that is a plane. We use coordinates $x_{\mu} = (x_a,y_i)$, with $x_a$ transverse to the plane and $y_i$ along the plane. (b) The transverse space to the plane.} 
\label{fig:plane}
\end{figure}

Throughout this paper, we consider an entangling surface $\Sigma$ that is a plane in flat space (see Fig.~\ref{fig:plane}).  The directions along $\Sigma$ are denoted by $y_i$ and the directions orthogonal to $\Sigma$ by $x^a$, so that $x^{\mu} = (x^a,y^i)$. Also,  $\Sigma$ is chosen to lie at the origin of the transverse space, $(x_1,x_2) = 0$. The region $V$ is thus the half-space, $x_1>0$. Furthermore, the transverse space to $\Sigma$ is $\text{O}(2)$ invariant, with an associated Killing field  $\xi=x_1\del_2-x_2\del_1$. The generator of rotations is the analytic continuation of the Rindler Hamiltonian,
\be \label{eq:HR}
H_R = - \int_A {T_{\mu \nu} \xi^{\mu} n^{\nu} } ~,
\ee
where $A$ is a constant Euclidean Rindler time slice (one of constant $\tan \theta = x_2/x_1$), and $n^{\nu}$ is normal to $A$.

The path integral defining the reduced density matrix can be interpreted in terms of angular evolution of the state at $\theta=0$ to the state at $\theta=2\pi$, with the Rindler Hamiltonian being the generator of infinitesimal angular translations. This leads to the immediate conclusion that the modular and Rindler Hamiltonians are proportional \cite{KabStra},
\be \label{eq:Kplane}
K = 2\pi H_R ~.
\ee
For instance, if one takes $A$ to lie on the $\theta=0$ slice,
\be \label{eq:Aplane}
A= \{(x_1,x_2,y_i) \in \mathcal{R}^d \, \big| x_2=0, x_1>0\}~,
\ee
 the modular Hamiltonian takes the form
\begin{equation} \label{eq:Kplane2}
K = -2\pi\int_{\Sigma} \int_0^\infty dx_1\, x_1\, T_{22} ~.
\end{equation}

We will consider a theory which contains a relevant or marginal operator $\mathcal{O}(x)$ of scaling dimension $\Delta\leq d$. The action $I$ thus has the term 
\be \label{eq:action}
 \lambda\int_{\mathbb{R}^d} \mathcal{O}(x) \ \subset I~.
\ee
We would like to find the change in the entanglement entropy resulting from a change $ \delta \lambda$ in the coupling. To accomplish this, we first note that the energy-momentum tensor of the full theory can be decomposed as\footnote{Throughout this paper we assume that $\mathcal{O}$ is independent of the background metric, \eg  in a gauge theory we exclude from consideration operators like $F_{\mu\nu}F^{\mu\nu}$.\label{Orestrict}}
\be \label{eq:Tmunu}
 T_{\mu\nu}=T_{\mu\nu}^{0}-\delta_{\mu\nu}\,\lambda\,\mathcal{O}~,
\ee
where $T_{\mu \nu}^{0}$ is the energy-momentum tensor of the theory with $\lambda =0$. Thus, from (\ref{eq:Kplane2}) we find that the change in the modular Hamiltonian is 
\be
K_{\lambda} - K_{\lambda_0} = 2 \pi\ \delta \lambda \int_{A}{x_1\ \mathcal{O}(x)},
\ee
Now, from the first law of entanglement entropy (\ref{eq:firstlaw}), we find that the change in entanglement entropy, to first order in $\delta \lambda$, is
\be \label{eq:deltaS1}
\delta S = -2\pi\ \delta \lambda\int_{A} x_1  \langle 0|\mathcal{O}(x) K_{\lambda_0}|0\rangle_{\lambda_0}~.
\ee
Since we could have evaluated the modular Hamiltonian on any constant Rindler time slice, and not necessarily the $\theta=0$ slice, we can use $\text{O}(2)$ symmetry in the $(x_1,x_2)$ space to alternatively write (\ref{eq:deltaS1}) in the form
\be \label{eq:deltaS2} 
\delta S = - \delta \lambda \int_{\mathbb{R}^d}{\langle 0| \mathcal{O}(x) K_{\lambda_0} |0 \rangle~}_{\lambda_0}.
\ee
This result matches the expression  found in \cite{RS2,RS1}. There, we considered an arbitrary entangling surface in an arbitrary background  and made use of the path integral representation of the reduced density matrix to find the appropriate generalization of (\ref{eq:deltaS2}). Here, we have worked with a plane in flat space. This is more restrictive,  but has the advantage that the modular Hamiltonian is known for any QFT and so we can directly use \reef{eq:firstlaw} to find (\ref{eq:deltaS1}). In the more general case \cite{RS2}, we instead did a perturbative expansion within the path integral to obtain (\ref{eq:deltaS2}).

\subsection{First order}
\label{1st}
As an illustration of our result (\ref{eq:deltaS1}), we use it to find the universal part of entanglement entropy for a free massive scalar field in four space-time dimensions.  In the notation of (\ref{eq:action}), the mass operator is $\mathcal{O}=\phi^2$ and the coupling constant is $\lambda=m^2/2$. Here, we view the mass term as being a small deformation of the massless theory, so $\delta \lambda = m^2/2$.~\footnote{Strictly speaking $\lambda$ is a dimensionful parameter, and so one needs to be more precise about the meaning of `a small deformation'. In general, it means $\delta\lambda\leq\lambda$. If, however, $\lambda$ vanishes, then by dimensional analysis we simply evaluate those  terms that may contribute to the universal entanglement entropy.} Later, in Sec.~\ref{sec:nonlinear}, we consider the more general case of an arbitrary  mass term which is regarded as a deformation of a theory with a slightly different mass. 

From (\ref{eq:deltaS1}) we find that the first order in $m^2$ contribution to the entanglement entropy is
\be \label{eq:deltaSm}
\delta S = - 2\pi^2 m^2 \int_{\Sigma}\int_0^{\infty}dx_1\, x_1\, \int_{\Sigma}\int_{-\infty}^0d\bar{x}_1\, \bar{x}_1\, \langle \mathcal{O}(x)\ T_{22}^0(\bar{x})\rangle_0~.
\ee
We are free to take the Rindler Hamiltonian to be evaluated on any constant Rindler time slice, and we have chosen $\theta = \pi$. 

The correlation functions we need to evaluate in (\ref{eq:deltaSm}) are for a free massless scalar field theory. The two-point function of a massless scalar is
\begin{equation} \label{eq:2point}
\langle \phi(x) \phi(0) \rangle = \frac{1}{(d-2) \Omega_{d}} \frac{1}{x^{d-2}}~,
\end{equation}
where $\Omega_d$ is the solid angle, $\Omega_d = 2 \pi^{d/2}/ \Gamma(d/2)$. The energy-momentum tensor for the minimally coupled scalar is given by
\begin{equation} \label{eq:Tscalar}
T_{\mu \nu}^{0} = \partial_{\mu} \phi \partial_{\nu} \phi - \frac{1}{2} \delta_{\mu \nu} (\partial \phi)^2 ~.%- \xi(\delta_{\mu \nu} \partial^2 - \partial_{\mu} \partial_{\nu})\phi^2~.
\end{equation}
Using this expression and the two-point function (\ref{eq:2point}) it follows that 
\begin{equation} \label{eq:TO}
\langle T_{\mu\nu}^{0}(\bar{x})\, \phi^2(x)\rangle = \frac{2(x_\mu - \bar{x}_\mu)(x_\nu - \bar{x}_\nu) - \delta_{\mu\nu}(x-\bar{x})^2 }{\Omega_d^2\, (x-\bar{x})^{2d}}~.
\end{equation}
Equipped with the correlator (\ref{eq:TO}), from (\ref{eq:deltaSm}) we find the first order change in entanglement entropy is,
\begin{equation} \label{eq:deltaSscalar0}
\delta S=  \frac{(2 \pi)^2 m^2}{2 \Omega_d^2}\int{d^{d-2}y}\int_0^{\infty}dx_1\, x_1 \int{d^{d-2}}\bar{y} \int_{-\infty}^{0} d \bar{x}_1\, \bar{x}_1  \frac{1}{((x_1-\bar{x}_1)^2 +(y - \bar{y})^2)^{d -1}}~.
\end{equation}
Performing the integral over $\bar{y}$ through a change of variables $\bar{y} \rightarrow y + \bar{y}$  yields
\be \label{eq:deltaSscalar1}
\delta S = \frac{m^2 \pi^{\frac{d+3}{2}} }{2^{d-3}\ \Gamma\(\frac{d-1}{2}\) \Omega_d^2}\int{d^{d-2}y}\int_0^{\infty}dx_1\, \int_{-\infty}^{0} d \bar{x}_1\, \frac{x_1\ \bar{x}_1}{(x_1-\bar{x}_1)^{d}}~.
\ee
In dimensions greater than four, $\delta S$ diverges as $\delta^{- (d-4)}$, where $\delta$ is the UV cutoff. In $4$ dimensions (\ref{eq:deltaSscalar1}) gives a log divergence,
\begin{equation} \label{eq:deltaSscalar2}
\delta S  = \frac{m^2}{24 \pi} \log (m \delta) \mathcal{A}_{\Sigma}~,
\end{equation}
where $\mathcal{A}_{\Sigma}$ is the area of the entangling surface. The result (\ref{eq:deltaSscalar2}) for the minimally coupled scalar matches the result in the literature (for example, \cite{Hertzberg:2010uv,Lewkowycz:2012qr}). 

As shown in \cite{MRS}, if  we had chosen the scalar field to be non-minimally coupled, the coefficient in (\ref{eq:deltaSscalar2}) would have changed. Indeed, from the general analysis it follows that $\langle T_{\mu \nu} \phi^2 \rangle$ vanishes for any CFT, and as a result if one deforms away from the fixed point, nontrivial contributions to the universal part of entanglement entropy will only appear at second order in the coupling \cite{RS1,Hung:2011ta}. The distinction between the entanglement entropies for the minimally and nonminimally coupled scalar field, and its implications, is elaborated and discussed in \cite{MRS}.

\subsection{All orders} \label{sec:allorders}
In the previous section, we considered the case of a free scalar field theory. In particular, we used \reef{eq:deltaS2} to find the first order change in the entanglement entropy under a variation in the mass of the field. In this section, we consider a general QFT and expand the entanglement entropy to all orders in $\delta \lambda$.

 First, we note that (\ref{eq:deltaS2}), as it is valid for any $\lambda_0$, leads to the exact differential equation \cite{RS2},
\be \label{eq:Sall1}
\frac{\partial S}{\partial \lambda} = - \int d^d x\, \langle  K_{\lambda}\, \mathcal{O}(x) \rangle~_{\lambda}~.
\ee
Here and in what follows we assume that $\lambda$ is a renormalized coupling constant, and therefore $\mathcal{O}(x)$ is the corresponding renormalized composite operator\footnote{Renormalization of the reduced density matrix is analogous to renormalization of  the generating functional in the field theory. Both of them have a path integral representation, with the only difference being in the boundary conditions, which don't have an impact on the renormalization procedure.}. Note also that $K_{\lambda}$ needs no renormalization since by definition the energy-momentum tensor is a finite composite operator. 

Using that $K_{\lambda}$ is the Rindler Hamiltonian (\ref{eq:Kplane}), we rewrite (\ref{eq:Sall1}) as
\be \label{eq:Sall2}
\frac{\partial S}{\partial \lambda} = 2 \pi\, \int d^d x\,\int_{0}^{\infty}d \bar{x}_1\, \bar{x}_1\, \int_{\Sigma}\  \langle  T_{22}^{\lambda}(\bar{x})\, \mathcal{O}(x) \rangle~_{\lambda}~.
\ee
We now expand both sides of (\ref{eq:Sall2}) about $\lambda_0$. Expanding the left hand side gives
\be
\frac{\partial S}{\partial \lambda} = \left(\frac{\partial S}{\partial \lambda}\right)\Big|_{\lambda=\lambda_0} + \left(\frac{\partial^2 S}{\partial \lambda^2}\right)\Big|_{\lambda=\lambda_0} \delta \lambda + O(\delta\lambda^2)~.
\ee
We expand the right side of (\ref{eq:Sall2}) by introducing a source $J(x)$ for $\mathcal{O}$ and taking functional derivatives with respect to it,
\begin{multline} \label{eq:TO1}
 \langle T_{\al \bt}^{\lambda} (\bar{x}) \mathcal{O}(x) \rangle_\lambda=\langle  T_{\al\bt}^{\lambda_0}(\bar{x}) \mathcal{O}(x) \rangle_{\lambda_0}+  \delta\lambda \int d^{d}z \,
 \, {\delta\over \delta J(z)} \langle T_{\al\bt}^{J}(\bar{x}) \mathcal{O}(x) \rangle_{J}\Big|_{\lambda=\lambda_0}+ O(\delta\lambda^2) ~.
\end{multline}
Since the energy-momentum tensor has the form (\ref{eq:Tmunu}),
functionally differentiating it gives
\be
{\delta T_{\al\bt}^{J}(\bar{x}) \over \delta J(z)}= - \delta_{\al\bt}\, \delta(z-\bar{x})\mathcal{O}(\bar{x})~.
\ee
Thus, Eq.~\ref{eq:TO1} becomes
\begin{multline} \label{eq:TO2}
 \langle T_{\al \bt}^{\lambda} (\bar{x}) \mathcal{O}(x) \rangle_\lambda=\langle  T_{\al\bt}^{\lambda_0}(\bar{x}) \mathcal{O}(x) \rangle_{\lambda_0}\\
 -\delta\lambda \left( \delta_{\al \bt} \langle \mathcal{O}(\bar{x}) \mathcal{O}(x)\rangle_{\lambda_0} + \int d^{d}z \,
 \,  \langle T_{\al\bt}^{\lambda_0}(\bar{x}) \mathcal{O}(x) \mathcal{O}(z) \rangle_{\lambda_0} \right) +  O(\delta\lambda^2) ~,
\end{multline}
where the first term in parenthesis comes from the functional derivative of $T_{\al \bt}^{\lambda}$, while the second term comes from the source term in the full action. 
In terms of the modular Hamiltonian we get, 
\begin{multline}
\langle K_{\lambda} \mathcal{O}(x)\rangle_{\lambda} = \langle K_{\lambda_0} \mathcal{O}(x) \rangle_{\lambda_0} - \delta \lambda \int d^d z\, \Big(\langle K_{\lambda_0} \mathcal{O}(x) \mathcal{O}(z)\rangle_{\lambda_0} - \langle \mathcal{O}(z) \mathcal{O}(x)\rangle_{\lambda_0}\Big) +O(\delta \lambda^2)~.
\end{multline}
Here we have made use of the fact that the modular Hamiltonian, $K_\lambda$, on the left hand side is invariant under $\text{O}(2)$ rotations in the transverse space, and therefore the whole correlator is some function of the distance from the entangling surface. In particular, we can replace
\be
2\pi\int_{\Sigma} \int_0^{\infty} d \bar{x}_1 \bar{x}_1 \langle \mathcal{O}(\bar{x})\mathcal{O}(x)\rangle = \int d^dz \langle \mathcal{O}(z)\mathcal{O}(x)\rangle~.
\ee
Now, matching terms in (\ref{eq:Sall2}) between the expansions of the left and the right hand sides in powers of $\delta \lambda$, gives the desired expansion of the entanglement entropy
\be \label{eq:Sall}
S(\lambda) = \sum_{n=0}^\infty{\frac{(\delta\lambda)^n}{n!}  \left(\frac{\partial^n S}{\partial \lambda^n}\right)\Big|_{\lambda_0}}~,
\ee
where the first-order term is
\be
\left(\frac{\partial S}{\partial \lambda}\right)\Big|_{\lambda_0} = -\int d^d x \langle K_{\lambda_0} \mathcal{O}(x) \rangle_{\lambda_0}~,
\ee
the second-order term is
\be
\left(\frac{\partial^2 S}{\partial \lambda^2}\right)\Big|_{\lambda_0} =\int d^d x\, \int d^d z\, \Big(\langle K_{\lambda_0} \mathcal{O}(x) \mathcal{O}(z)\rangle_{\lambda_0} - \langle \mathcal{O}(z) \mathcal{O}(x)\rangle_{\lambda_0}\Big)~,
\ee
and similarly the $n$-th order term is 
\be
\left(\frac{\partial^n S}{\partial \lambda^n}\right)\Big|_{\lambda_0} =(-1)^{n} \underset{n}{\underbrace{~\idotsint}} 
\Big(\langle K_{\lambda_0} \underset{n}{\underbrace{\mathcal{O} \cdots \mathcal{O}}}\rangle_{\lambda_0} 
- (n-1) \langle \underset{n}{\underbrace{\mathcal{O} \cdots \mathcal{O}}} \rangle_{\lambda_0}\Big)~.
\label{nth-order}
\ee

Although the above expression looks universal for small $n$, it is in fact theory dependent. The information of the underlying field theory is hidden in the structure of the renormalized composite operators $\mathcal{O}$. Generically this structure can be very complicated. Moreover, even if $\lambda_0$ corresponds to a fixed point where all anomalous dimensions vanish, one should still anticipate emergence of a non-universal structure at sufficiently high order in $\delta\lambda$.

As a side note, we mention that in deriving \reef{nth-order} we have ignored operator ordering. It is implicit that when applying \reef{nth-order} one may need to include contact terms. It is interesting to note  that an alternative way to obtain \reef{nth-order} would be to do an expansion of the definition of $S$, (\ref{eq:Sdef}), to all orders in $\delta \rho$. This would appear to be a more challenging approach, as one would have to worry about issues relating to potential non-commutativity of $K_{\lambda_0}$ with $\delta \rho$ \cite{Blanco:2013joa}. Since the commutators are of local operators and nonzero only at coincident points, if one ignores this issue and just does a naive Taylor expansion, one recovers \reef{nth-order}. Though of course, our derivation  of \reef{nth-order} was special to a planar entangling surface.

\section{Flat entangling surface: non-perturbative analysis} \label{sec:nonlinear}
In the previous section we found a differential equation (\ref{eq:Sall1}) that encodes the dependence of the entanglement entropy $S$ on the coupling $\lambda$ of some relevant operator $\mathcal{O}$. Expressed in terms of the modular Hamiltonian for a plane, we found,
\be \label{eq:Sall2v2}
\frac{\partial S}{\partial \lambda} = 2 \pi\, \int d^d x\,\int_{0}^{\infty}d \bar{x}_1\, \bar{x}_1\, \int_{\Sigma}\  \langle  T_{22}^{\lambda}(\bar{x})\, \mathcal{O}(x) \rangle~_{\lambda}~.
\ee
Provided that one knows the two-point function of the relevant operator and the energy-momentum tensor, $\langle T_{\mu \nu} \mathcal{O} \rangle$, the problem of computing entanglement entropy for a plane is solved by integrating (\ref{eq:Sall2v2}). Of course, one generally only knows such a correlation function perturbatively. For this reason, in Sec.~\ref{sec:allorders} we derived a perturbative expansion of $S$ around some given coupling $\lambda_0$ in terms of the unperturbed correlation functions defined at $\lambda=\lambda_0$.

For a few special cases, one can exactly evaluate the right hand side of (\ref{eq:Sall2v2}). The simplest case is that of a free massive scalar or fermion, and this is what we do in Sections \ref{sec:fermion}-\ref{sec:min}. From a theoretical standpoint, it is of interest to evaluate (\ref{eq:Sall2v2}) for a general theory, expressing the entanglement entropy in terms of some parameters characterizing the theory. This is done in Sec.~\ref{sec:gen}, for theories with one relevant operator, by making use of the spectral function.

\subsection{General case} \label{sec:gen}
We start with the trace Ward identity \cite{Osborn:1993cr,Jack:2013sha}, 
\be \label{eq:Ward}
\langle T(x) \rangle_{\lambda} + \sum_{i}(d-\Delta_i+ \beta_i)\, \lambda_i\, \langle \mathcal{O}_i(x) \rangle_{\lambda}=\mathcal{A}~,
\ee
where $\mathcal{O}_i$ are all relevant operators and $\mathcal{A}$ is the trace anomaly. Differentiating (\ref{eq:Ward}) with respect to $g^{\al\bt}(y)$ gives \footnote{See footnote \ref{Orestrict}.},
\bea  \label{eq:trace2}
 \langle T(x) T_{\al\bt}(y) \rangle_\lambda + \sum_{i}(d-\Delta_i+\beta_i)\, \lambda_i\, \langle \mathcal{O}(x) T_{\al\bt}(y) \rangle_\lambda 
 \non
 = {2\over\sqrt{g(y)}} \left\langle{\delta  T(x) \over \delta g^{\al\bt}(y)}\right\rangle_\lambda - {2\over\sqrt{g(y)}} {\delta\mathcal{A}\over\delta g^{\al\bt}(y)} ~.%\\
\eea
Since the couplings $\lambda_i$ are constant and $g_{\mu\nu}$ is flat, the right hand side of (\ref{eq:trace2}) reduces to a $\delta$-function with constant coefficient. Thus, the terms on the right hand side do not contribute to the universal part of entanglement entropy, and we drop them in what follows. Now using (\ref{eq:trace2}) combined with (\ref{eq:Sall2v2}) gives
\be  \label{eq:SallTT}
  \sum_{i}{\del S \over\del \lambda_i} \lambda_i\, (d-\Delta_i + \beta_i)=-{2\pi} \int d^d x \int_{\Sigma}\int_0^\infty d\bar{x}_1\ \bar{x}_1 \langle \, T(x) \, T_{22}(\bar{x})\rangle_{\lambda} ~.
\ee
As the right hand side can be expressed in terms of the modular Hamiltonian we get,
\be  \label{eq:SallTT2}
  \sum_{i}{\del S \over\del \lambda_i} \lambda_i\, (d-\Delta_i + \beta_i)= \int d^d x \, \langle \, T(x) \, K_{\lambda}\, \rangle_{\lambda} ~.
\ee

As an aside, we mention that in Sec.~\ref{sec:defgeom} we will see that the expression on the right hand side is what appears when considering the variation of the entanglement entropy under a change of the spacetime metric (rather than the coupling). In addition, although we arrived at (\ref{eq:SallTT2}) for a planar entangling surface, it in fact holds for any entangling surface in any spacetime~\cite{RS2}, although  generally $K_{\lambda}$ is unknown.

Returning to (\ref{eq:SallTT2}), we see that it is especially useful if the theory has  only one relevant operator. For the rest of the section we will therefore specialize to this case.  A feature of (\ref{eq:SallTT}) is that there exists a spectral decomposition of the two-point function of the energy-momentum tensor \cite{Cappelli:1990yc}, allowing it to be expressed in terms of a sum involving free propagators, 
\be
G(x-\bar{x}, \mu)=\int {d^dp \over (2\pi)^d} {e^{ip\cdot(x-\bar{x})}\over p^2+\mu^2} = \frac{1}{2\pi}\left(\frac{\mu}{2\pi |x-\bar{x}|}\right)^{(d-2)/2} K_{(d-2)/2}(\mu |x-\bar{x}|) ~,
\ee
and two spectral functions: $c^{(0)}(\mu)$ and $c^{(2)}(\mu)$. The spectral decomposition takes the form
\bea
 \langle  T_{\al\bt}(x)  T_{\rho\sigma}(\bar{x}) \rangle &=& {A_d \over (d-1)^2} \int_0^\infty d \mu ~ c^{(0)}(\mu) ~ \Pi^{(0)}_{\al\bt,\rho\sigma}(\del)~G(x-\bar{x},\mu) 
 \non
 &+&{A_d \over (d-1)^2} \int_0^\infty d\mu~ c^{(2)}(\mu) ~ \Pi^{(2)}_{\al\bt,\rho\sigma}(\del)~G(x-\bar{x},\mu) 
  ~,
 \label{masscorr}
\eea
where 
\bea
 A_d&=&{\Omega_d\over (d+1) 2^{d-1}}~, \quad \Omega_d={2\pi^{d/2}\over \Gamma\(d/2\)} ~,
 \non
 \Pi^{(0)}_{\al\bt,\rho\sigma}(\del)&=& {1\over \Gamma(d)} S_{\al\bt} S_{\rho\sigma}\, ,
 \non
 \Pi^{(2)}_{\al\bt,\rho\sigma}(\del)&=& {d-1\over 2\,\Gamma(d-1)}\( S_{\al\rho}S_{\bt\sigma}+S_{\al\sigma}S_{\bt\rho} - {2\over d-1}S_{\al\bt} S_{\rho\sigma} \)~,
\eea
where $S_{\al\bt}=\del_\al\del_\bt- \delta_{\al\bt}\del^2$. 

Using the above spectral representation of the two point function combined with the modular Hamiltonian (\ref{eq:Kplane2}), one finds that for indicies $i,j$ along the entangling surface $\Sigma$ \cite{conifold},
\bea
 \langle T_{ij}(r)  K_\lambda \rangle&=& -{A_d \, \delta_{ij}\over (d-1)^2 \Gamma(d)} \int_0^\infty d\mu ~ \big(c^{(0)}(\mu) -(d-1)c^{(2)}(\mu) \big)
 \, \mu^2 K_0(\mu\, r)~,
 \non
 \label{TijH}
\eea
where $r=\sqrt{ x_2^2+ x_1^2}$ is the radial distance in the transverse space (the distance between $\Sigma$ located at the origin $r=0$, and the insertion point of the energy-momentum tensor). On the other hand, for indicies transverse to $\Sigma$ we have,
\begin{multline}
 \langle T_{ac}(r)  K_\lambda \rangle  = -{A_d \over (d-1)^2 \Gamma(d)} \int_0^\infty d\mu ~ \big(c^{(0)}(\mu) +(d-1)(d-2)c^{(2)}(\mu) \big)
 \\
 \times(\delta_{ac}\mu^2 -\del_a\del_c)  K_0(\mu r )~.
  \label{TacH}
\end{multline}
Now using the Bessel equation it can be shown that,
\be
 \del_a\del_c  K_0(\mu\, r) ={\mu \over r} \left(\begin{array}{cc} \mu r  K_2(\mu r)\cos^2\theta -  K_1(\mu r) & \mu r\sin\theta\cos\theta \, K_2(\mu r) 
 \\ \mu r \sin\theta\cos\theta \,K_2(\mu r) & \mu r  K_2(\mu r)\sin^2\theta -  K_1(\mu r)\end{array}\right)~,
\ee
where we substituted $ x_1=r\cos\theta$ and $ x_2=r\sin\theta$. Hence, taking the trace and using (\ref{TacH}) and (\ref{TijH}) yields,
\be \label{blah}
 \langle \,  T(r) \,  K_\lambda \,\rangle_\lambda={ - A_d \over (d-1) \Gamma(d)}  \int_0^\infty d \mu ~ \mu^2 \, c^{(0)}(\mu) \, K_0(\mu\, r)~.
\ee
Note that $c^{(2)}(\mu)$ does not contribute since by definition $\Pi^{(2)}_{\al\bt,\rho\sigma}$ is manifestly traceless in the first and last pair of indices. 

Finally, substituting the above result (\ref{blah}) into (\ref{eq:SallTT2}) and integrating over $\mathbb{R}^d$ gives
\be
  \lambda\,{\del S \over\del \lambda}=
  { - 2\pi\,A_d\,\A_\Sigma \over (d-\Delta+\beta_\lambda)(d-1)\Gamma(d)}  \int_0^\infty d\mu ~ \, c^{(0)}(\mu) ~,
  \label{CS3}
\ee
where $\A_\Sigma$ is the area of the entangling surface $\Sigma$. Of course, for a conformal field theory (for dimensions greater than $2$) the above identity is trivial since both $\lambda$ and $c^{(0)}(\mu)\propto \mu^{d-2}\delta(\mu)$  vanish. Interestingly, the right hand side of (\ref{CS3}) exhibits a logarithmic divergence if and only if the expansion of $c^{(0)}(\mu)$ for large values of $\mu$ contains a term that behaves as $\mu^{-1}$. 

Finding the entanglement entropy is now a simple matter of inserting the relevant spectral function into (\ref{CS3}). We do this now for the simplest case of free massive fields. 

\subsection{Dirac fermion} \label{sec:fermion}

For a free massive Dirac field, $\lambda=m$, $\Delta=d-1$, $\bt_\lambda=0$, and the spectral functions are \cite{Cappelli:1990yc} (see also Appendix \ref{spectr}),
\bea
 c^{(0)}_F(\mu)&=&2^{[d/2]}\, {2(d+1)(d-1)\over \Omega_d^2} \, m^2 \, \mu^{d-5}\(1-{4m^2\over \mu^2}\)^{(d-1)/2}\Theta(\mu-2m)~,
 \label{cF}
 \\
 c^{(2)}_F(\mu)&=&2^{[d/2]}\, {(d-1)\over 2 \, \Omega_d^2} \, \mu^{d-3}\(1-{4m^2\over \mu^2}\)^{(d-1)/2}\(1+{2\over d-1}{4m^2\over \mu^2}\)\Theta(\mu-2m)~,
 \nonumber
\eea
where $m$ is the mass of the Dirac field and $[d/2]$ denotes the integer part of $d/2$. (Recall that $2^{[d/2]}$ is the dimension of the Clifford algebra in a $d$-dimensional spacetime.) Although we will not need it here, we have recorded $c^{(2)}$ as it will be relevant in Sec.~\ref{sec:defgeom}. Using $c^{(0)}_F$, (\ref{CS3}) takes the form 
\begin{multline}
  m \,{\del S \over\del m}=
  { - 8\pi\over 2^{d/2} \Gamma(d)\Omega_d} \, m^{d-2}\A_\Sigma
  \int_{2}^{\infty} dx \, x^{d-5} \(1-{4\over x^2}\)^{(d-1)/2} 
   = - {\Gamma\({4-d\over 2}\)\over 6 (2\pi)^{d-2\over 2}} \, m^{d-2}\A_\Sigma
  \\
  ={(-)^{d/2} \over 6(2\pi)^{d-2\over 2}\Gamma(d/2)} \,(d-2)\, m^{d-2}\A_\Sigma~ \log(m\delta)+\ldots~,
  \label{fermCS}
\end{multline}
where by assumption $d$ is even, $\delta$ is the UV cut-off, and we have kept only the logarithmic divergence\footnote{The logarithmic divergence corresponds to a pole in the gamma function. The precise correspondence is $\epsilon^{-1}=\log(m\delta)$, where $d=2n+\epsilon$ and $n$ is an integer. This correspondence can be read off by introducing a sharp cut-off, $(m\delta)^{-1}$, in the upper limit of the integral in (\ref{fermCS}).\label{divgnce}} since this is the only part that is relevant for (\ref{CS3}). We thus recover the universal `area' term for fermions \cite{Hertzberg:2010uv}
\be \label{eq:SfermF}
 S ={(-)^{d/2} \over 6(2\pi)^{d-2\over 2}\Gamma(d/2)} m^{d-2}\A_\Sigma\, \log(m \delta)~.
\ee

\subsection{Scalar} \label{sec:min}
In  Appendix \ref{spectr} we find the spectral function $c^{(0)}_M$ for the minimally coupled scalar,
\be
c^{(0)}_M(\mu)=\mu^{d-3}{8(d-1)(d+1) \over \Omega_d^2} \(1-4{m^2\over \mu^2}\)^{d-3\over 2}\({d-2\over 4}+ {m^2\over\mu^2}\)^2\Theta(\mu-2m)~.
\ee
Hence, we obtain
\begin{multline}
  m^2\,{\del S\over\del m^2}=
  { - 16\pi\over 2^d \, \Gamma(d)\Omega_d} \, m^{d-2}\A_\Sigma
  \int_{2}^{\infty} dx \, x^{d-3} \(1-{4\over x^2}\)^{(d-3)/2} \( {d-2\over 2} + {1\over x^2}  \)^2
   \\
   =  - {\Gamma\({4-d\over 2}\)\over 12 (4\pi)^{d-2\over 2}} \, m^{d-2}\A_\Sigma~.
\end{multline}
Expanding around even $d$ gives
\be \label{eq:SscalarF}
  S=   {(-)^{d\over 2}\over 6 (4\pi)^{d-2\over 2}\Gamma(d/2)} 
  \, m^{d-2}\A_\Sigma \, \log(m\delta)  ~.
  \ee
Thus we recover the universal `area'' term for the minimally coupled scalar \cite{Hertzberg:2010uv, Huerta:2011qi, Lewkowycz:2012qr}. 

Remarkably, the result for the entanglement entropy is sensitive to the way the scalar field couples to gravity. For the nonminimally coupled scalar one can regard the additional contributions to the entanglement entropy as arising from a boundary term in the modular Hamiltonian.  We thoroughly investigated the general case (and in particular the conformally coupled scalar field) in \cite{MRS}.

\section{Deformed geometry: new universal terms} \label{sec:defgeom}
In the previous sections we studied the dependence of entanglement entropy on the coupling constants of the operators in the theory (see Eq.~\ref{eq:Sall1}). One could also ask how entanglement entropy depends on the geometry. This is encoded in the analogous flow equation to (\ref{eq:Sall1}) \cite{RS2,RS1}
\be \label{eq:deltaSgeom}
 \frac{\delta S}{\delta g^{\mu \nu}(x)} =-\frac{1}{2} \sqrt{g}\ \langle T_{\mu \nu}(x) K_{\lambda} \rangle_{\lambda} ~,
\ee
where $\langle \ldots \rangle_{\lambda}$ is the connected two-point function in the vacuum state of the theory with coupling $\lambda$, and $K_{\lambda}$ is the modular Hamiltonian for the entangling surface $\Sigma$ under consideration.

The equation (\ref{eq:deltaSgeom}) is completely general, in that it holds for any background geometry and any entangling surface. However, solving (\ref{eq:deltaSgeom}) is challenging, as we only know the modular Hamiltonian $K_{\lambda}$ for a planar entangling surface in flat space. We must therefore resort to solving (\ref{eq:deltaSgeom}) perturbatively. In this section we will consider only the first order piece, 
\be \label{eq:varEE}
 \delta S={1 \over 2}\int d^{d} x \langle T^{\mu\nu}(x) K_\lambda \rangle_{\lambda} \, h_{\mu\nu} +\mathcal{O}(h^2)~,
\ee
where $h_{\mu\nu}$ encodes all the information about changes in the geometry. We will be considering perturbations in geometry around a planar entangling surface in flat space, so the modular Hamiltonian $K_{\lambda}$ is the Rindler Hamiltonian (\ref{eq:Kplane}) and is an integral of the energy-momentum tensor. The first order terms in the perturbation of the metric can be written as \cite{RS1}
\begin{eqnarray}
h_{ij} &=& x^{a}x^{c}\mathcal{R}_{i a c j} + \mathcal{O}(x^3)~, \non
h_{ab} &=& -\frac{1}{3} \mathcal{R}_{a c b d}x^c x^d  + \mathcal{O}(x^3)~,
\label{h}
\end{eqnarray}
where $\mathcal{R}_{\mu \nu \alpha \beta}$ is the Riemann tensor, and directions $i,j$ are along the entangling surface while directions $a,b$ are transverse to it. 

In \cite{RS1} we evaluated (\ref{eq:varEE}) for a CFT, reproducing the first order terms of the general expression for a $4$ dimensional CFT found in \cite{solo}. These terms are the purely geometrical contributions to the universal part of entanglement entropy. Here we want to consider terms in $S$ which involve mixing of geometry with the couplings of the theory. 

In Sec.~\ref{sec:genG}, we evaluate $\delta S$ given by (\ref{eq:varEE}) for a general QFT, expressing the answer in terms of the spectral functions of the theory. Then in Sec.~\ref{sec:geomfermion} and Sec.~\ref{sec:geomscalar} we evaluate $\delta S$ for the special case of massive free fields. While in Sec.~\ref{sec:nonlinear} we considered a planar entangling surface in flat space, which allowed us to identify ``area'' term contributions to entanglement entropy of the form $m^{d-2} \mathcal{A}_{\Sigma} \log (m \delta)$, the perturbed geometry in Sec.~\ref{sec:geomfermion},~\ref{sec:geomscalar} will allows us to identify contributions that involve mixing of the masses of free fields with the background curvature.

\subsection{General case} \label{sec:genG}
To evaluate the first order change in entanglement entropy under a deformation of the background geometry, we need only to evaluate (\ref{eq:varEE}). It will be convenient to express the answer in terms of the spectral functions appearing in the two-point function of the energy-momentum tensor that were introduced in \cite{Cappelli:1990yc} and reviewed in Sec.~\ref{sec:nonlinear}. By making use of the correlator $\langle T_{\mu \nu} K\rangle$ found in Sec.~\ref{sec:nonlinear} (equations (\ref{TijH}) and (\ref{TacH})), and combining it with the expression (\ref{h}) for $h_{\mu \nu}$, we get  $\delta S$ from (\ref{eq:varEE}),
\bea
\delta S &=&{2\pi A_d \over (d-1)^2 \Gamma(d)} \int_{\Sigma} \( \delta^{ac}\delta^{ij}R_{iajc} +{1\over 2}\delta^{ac}\delta^{bd}R_{abcd} \)
\int_0^\infty {d\mu \over \mu^2} c^{(0)}(\mu)
\non
 &+&{2\pi A_d \over (d-1) \Gamma(d)} \int_{\Sigma} \({d-2\over 2}\delta^{ac}\delta^{bd}R_{abcd} - \delta^{ac}\delta^{ij}R_{iajc}\)\int_0^\infty {d\mu\over \mu^2} c^{(2)}(\mu)~.
 \label{varEE2}
\eea

Using the Gauss-Codazzi relation one can show that for our choice of the unperturbed geometry, $\delta^{ij}\delta^{kl}R_{ikjl}$ is a total derivative \cite{RS1}. Hence, from the definition of the Weyl tensor we get,
\be
 \delta^{ac}\delta^{bd}C_{abcd}={d(d-3)\over (d-1)(d-2)}\Big( \delta^{ac}\delta^{bd}R_{abcd} - {2\over d} \delta^{ac}R_{ac} \Big)+\ldots~,
\ee
where ellipsis denote total derivatives and higher order terms in $h_{\mu\nu}$. Substituting this expression into (\ref{varEE2}), we finally obtain
\bea
\delta S &=&{-(d-2)\pi A_d \over (d-1)(d-3) \Gamma(d+1)} \int_{\Sigma} \( \delta^{ac}\delta^{bd}C_{abcd} - 2{d-3\over d-2}\delta^{ac}R_{ac} \)
\int_0^\infty {d\mu\over \mu^2} c^{(0)}(\mu)
\non
 &+&{ (d-2)\pi A_d \over (d-3) \Gamma(d)} \int_{\Sigma} \delta^{ac}\delta^{bd}C_{abcd} \int_0^\infty {d\mu \over \mu^2} c^{(2)}(\mu)~,
 \label{mixEE}
\eea

This formula is a general expression that describes mixing of the background curvature with the couplings of the theory. Of course, in general the spectral functions are not known. However, in certain cases, such as for a CFT or free field theories, they are known and we can evaluate $\delta S$ in closed form.

For instance, for the case of a CFT with central charge $c$, the spectral functions are \cite{Cappelli:1990yc}
 \be
 c^{(0)}(\mu)\Big|_{CFT} \propto \mu^{d-2}\delta(\mu) ~, \quad c^{(2)}(\mu)\Big|_{CFT}={d-1\over d} C_T\, \mu^{d-3}~,
\ee
where $C_T$ is the coefficient characterizing the leading singularity in the two-point function of two stress tensors \cite{Osborn:1993cr,Erdmenger:1996yc}. In four dimensions, this coefficient is related to the standard central charge $c$ which appears as the coefficient of the (Weyl)$^2$ term in the trace anomaly: $C_T=40\, c/\pi^4$. Substituting the CFT spectral functions into (\ref{mixEE}), we find that a logarithmically divergent term in $4$ dimensions is 
\be
\delta S = -\frac{c}{2\pi} \int_{\Sigma} \delta^{a c} \delta^{b d}\, C_{a b c d} \log(\delta)~,
\ee
in agreement with \cite{solo} and our findings in \cite{RS1}.
As is clear on dimensional grounds, for a CFT in dimensions higher than $4$, the terms in the expansion of the metric perturbation $h_{\mu \nu}$ which are explicitly shown in (\ref{h}) give rise to power law divergent contributions to the entanglement entropy. Fortunately, for a theory which is not a CFT, there are new scales which lead to $\log(\delta)$ terms in $d\geq 6$ without the need to consider other terms in the expansion of $h_{\mu \nu}$. We now turn to evaluating $\delta S$ for the free massive fermion and scalar.

\subsection{Dirac fermion} \label{sec:geomfermion}
For the free massive fermion, the spectral functions are given by (\ref{cF}). Substituting them into $\delta S$ given by (\ref{mixEE}) we obtain,
\be
\delta S=  {m^{d-4}\over 60 d } \,{\Gamma\({6-d\over 2}\)\over  (2\pi)^{d-2\over 2}}\,
  \( \int_{\Sigma} \delta^{ac}R_{ac} - 2 \, { (d-1)(d-2) \over (d-3)(d-4)} \,  \int_{\Sigma} \delta^{ac}\delta^{bd}C_{abcd}\)~.
\ee
Expanding this expression around even $d\geq 6$, yields
\be
\delta S=  {m^{d-4} \over 30 d } \,{(-)^{d/2}\over  (2\pi)^{d-2\over 2}\Gamma\({d-4\over 2}\)}\,
 \( \int_{\Sigma} \delta^{ac}R_{ac} - 2{ (d-1)(d-2) \over (d-3)(d-4)} \,  \int_{\Sigma} \delta^{ac}\delta^{bd}C_{abcd}\) \log(m\delta)~,
\ee
where the correspondence between cut-offs is $\epsilon^{-1}=\log(m\delta)$ with $d=2n+\epsilon$ and integer $n$.

\subsection{Scalar} \label{sec:geomscalar}
For the free massive conformally coupled scalar, the spectral functions are given by \cite{Cappelli:1990yc}
\bea \label{cS}
 c^{(0)}_S(\mu)&=&{8(d+1)(d-1)\over \Omega_d^2}\,m^4 \, \mu^{d-7}\(1-{4m^2\over \mu^2}\)^{(d-3)/2}\Theta(\mu-2m)~,
 \non
 c^{(2)}_S(\mu)&=& \mu^{d-3}\(1-{4m^2\over \mu^2}\)^{(d+1)/2}\Theta(\mu-2m)~.
\eea

Substituting them into (\ref{mixEE}) we obtain,
\be
\delta S=  { m^{d-4}\over 60 d(d-1) } \,{\Gamma\(4-{d\over 2}\)\over  (4\pi)^{d-2\over 2}}\,
 \left[  \, 
\int_{\Sigma} \delta^{ac}R_{ac} + {d-2\over 2(d-3)}\(\frac{2 d (d-1) \Omega_d^2}{(d-6)(d-4)} -1\) 
  \int_{\Sigma} \delta^{ac}\delta^{bd}C_{abcd}   \right].
\ee
In six dimensions only the Weyl tensor contributes to the universal entanglement entropy,
\be
 \delta S=  {\Omega_6^2 \over 90 (4\pi)^2 }\,m^2 \int_{\Sigma} \delta^{ac}\delta^{bd}C_{abcd}  ~ \log(m\delta)~,
\ee
whereas for even $d\geq 8$, we have
\begin{multline}
\delta S=  { m^{d-4} \over 30 d(d-1) } \,{(-)^{{d\over 2}+1}\over  (4\pi)^{d-2\over 2}\Gamma\({d-6\over 2}\) }
\\
\times
 \left[  \, 
\int_{\Sigma} \delta^{ac}R_{ac} + {d-2\over 2(d-3)}\(\frac{2 d (d-1) \Omega_d^2}{(d-6)(d-4)} -1\) 
  \int_{\Sigma} \delta^{ac}\delta^{bd}C_{abcd}   \right]\log(m\delta)~.
\end{multline}

\section{Discussion}

In this paper we exploited the fact that for a planar entangling surface in flat space, the modular Hamiltonian (defined as the $\log$ of the reduced density matrix) is known for any QFT. In Sec.~\ref{sec:defgeom} we made use of this to find  the entanglement entropy across a planar entangling surface in a background that is a slight deformation of flat space. Working to first order in the deformation of the metric, we found a new class of terms appearing in the universal part of entanglement entropy. These terms encode mixing of background curvature and the couplings of the theory. By making use of the spectral decomposition of the two-point function of the energy-momentum tensor, we gave a closed form answer for any QFT. Evaluating this part of entanglement entropy for some theory thus amounts to simply knowing the spectral functions. While for free massive field theories the spectral functions are known, it would be of obvious interest to consider interacting theories for which the spectral functions could be found perturbatively.

In Sec.~\ref{sec:nonlinear} we focused on relevant deformations of the field theory. The flow equation (\ref{eq:SIntro}) gives, in principle, a full expression for the dependence of entanglement entropy on all the couplings in the theory. For a general QFT, one needs to merely evaluate the two-point function $\langle T_{\mu \nu} \mathcal{O}_i\rangle$. We did this explicitly for free massive theories. It would be interesting to also consider interacting theories (such as $\lambda \phi^4$ \cite{Hertzberg:2012mn} or the $O(N)$ vector models \cite{Sachdev}), in which case one would make use of the perturbative expansion of the entanglement entropy given in Sec.~\ref{sec:allorders}. In the special case that there is only one relevant operator, we made use of the Ward identity to express the dependence of entanglement entropy on the coupling in terms of a two-point function of the energy-momentum tensor. A spectral decomposition of this two-point function allowed us to give a closed form expression for the entanglement entropy. It would be of theoretical interest to, if possible, directly do a spectral-like decomposition of $\langle T_{\mu \nu} \mathcal{O}_i\rangle$, and thereby obtain a closed form expression for the dependence of entanglement entropy on each of the couplings in the theory.

\acknowledgments  
We thank H.~Casini, R.~Myers, Y.~Nomura, M.~Rangamani, S.J.~Rey, and S.~Solodukhin for helpful discussions. This work is supported in part by NSF Grant PHY-1214644 and the Berkeley Center for Theoretical Physics. VR thanks the Aspen Center for Physics under NSF Grant 1066293 for hospitality during the completion of this work.

\appendix

\section{The thermal picture}

In this appendix we review how the entanglement entropy across a plane for a free field theory can be evaluated by using the entropy density of a free gas. Indeed, since an accelerated Unruh observer is confined to the right Rindler wedge, he traces out the degrees of freedom in the left Rindler wedge. Thus, entanglement entropy can be regarded as the entropy of Unruh radiation.  It is interesting that in fact one can easily find the entanglement entropy across a plane for a free massive theory by computing the entropy of Unruh radiation \cite{KabStra}.\footnote{To compute the entropy of the Unruh radiation one needs to impose some kind of boundary conditions at the entangling surface. It is not a priori obvious which boundary conditions are the correct ones to choose (Dirichlet are a popular choice \cite{'tHooft:1984re}), nor if the answer is sensitive to the choice. The choice we make is to use the entropy density of a thermal gas in the infinite volume limit.}

Recall that a uniformly accelerated observer experiences thermal radiation with temperature
\begin{equation} \label{eq:T}
T= \frac{1}{2\pi x}~,
\end{equation}
where $x$ is the transverse distance from the plane at $t=0$, and the observer's worldline is $x^2 - t^2 =a^2$. We will find the entanglement entropy by using standard thermodynamic formulas for the entropy density of a gas, making use of (\ref{eq:T}), and integrating over the entire right Rindler wedge. 

The entropy density will be found by differentiating the pressure density,
\begin{equation} \label{eq:s}
s= \frac{\partial p}{\partial T}~.
\end{equation}
The pressure is found by integrating the component $T_{ii}$ of the stress-tensor against the number density, 
\begin{equation} \label{eq:p1}
p = \frac{g}{(2\pi)^{d-1}} \int{d^{d-1} p\, \frac{p^2}{(d-1)\omega}\, \frac{1}{e^{\omega/T} \mp 1}}~,
\end{equation}
where $g$ is the degeneracy, and homogeneity has allowed us to replace $p_i^2$ with $p^2/(d-1)$. Using $\omega^2 = p^2 +m^2$, we rewrite (\ref{eq:p1}) as an integral over $\omega/T$,
\begin{equation} \label{eq:p2}
p = \frac{g}{d-1}\, \frac{\Omega_{d-1}}{(2\pi)^{d-1}}\, T^d\, f_{\mp}(\alpha)~,
\end{equation}
where we defined,
\begin{equation} \label{eq:f}
f_{\mp}(\alpha) =  \int_{\alpha}^{\infty}{du\, \frac{(u^2 - \alpha^2)^{\frac{d-1}{2}}}{e^u \mp 1}}~,
\end{equation}
and introduced the dimensionless parameter $\alpha \equiv m/T$. The leading order term in (\ref{eq:p2}) scales as $O(\alpha^0)$ and gives $p\sim T^d$. The entropy density is correspondingly, $s \sim T^{d-1}$. Integrating over the Rindler wedge and making use of (\ref{eq:T}) yields
\begin{equation}
S \sim \int{d^{d-2} y\ d x\ \frac{1}{x^{d-1}}} \sim \frac{\mathcal{A}_{\Sigma}}{\delta^{d-2}}~,
\end{equation}
which is the standard area law term. We are, however, interested in the $\log(\delta)$ term. This arises from the term in (\ref{eq:f}) that is of order $O(\alpha^{d-2})$. Performing a binomial expansion of the numerator in the integrand of (\ref{eq:f}), the relevant term is present in even space-time dimensions,
\begin{equation} \label{eq:f2}
f_{\mp}(\alpha) =-\alpha^{d-2} (-1)^{d/2}\frac{\Gamma(\frac{d+1}{2})}{\Gamma(d/2) \Gamma(3/2)}\int_0^{\infty}{d u \frac{u}{e^{u}\mp 1}} +...
\end{equation}
We find the entropy by integrating the entropy density (\ref{eq:s}), making use (\ref{eq:f2}), (\ref{eq:p2}), and (\ref{eq:T}). For bosons we use $f_{-}(\alpha)$ defined by (\ref{eq:f2}) and set $g=1$ to get
\begin{equation} \label{eq:Sbos}
S= -\frac{2 \pi (-1)^{d/2} }{6 (4 \pi)^{\frac{d-2}{2}} \Gamma(d/2)} m^{d-2}\int{d^{d-2} y} \int_{\delta}^{\infty}{\frac{1}{2\pi x}}~.
\end{equation}
The integral over the direction $x$ orthogonal to the entangling surface produces the logarithm of the UV cutoff, while the integral over the coordinates $y$ along the entangling surface gives the area $\mathcal{A}_{\Sigma}$. Thus, (\ref{eq:Sbos}) gives the standard result for the universal part of entanglement entropy for the minimally coupled scalar, (\ref{eq:SscalarF}). 
For fermions, we us $f_{+}(\alpha)$ defined by (\ref{eq:f2}) and degeneracy $g=2^{d/2}$ to find that the entanglement entropy matches (\ref{eq:SfermF}).

\section{Spectral density for free fields}
\label{spectr}

In this Appendix we show how to derive the spectral density, $c^{(0)}(\mu)$, in the case of free fields. In general, this spectral function is given by \cite{Cappelli:1990yc}
\be
 c^{(0)}(\mu)={2\,\Gamma(d)\over \pi A_d} \, {1\over \mu^3} \, \text{Im} \langle T(p) T(-p) \rangle \big|_{p^2=-\mu^2}~,
 \label{c0}
\ee
where the imaginary part of a given function $f(p^2)$ is defined by
\be
  \text{Im} f(p^2)|_{p^2=-\mu^2}=\lim_{\epsilon\to0} {f(-\mu^2-i\epsilon)-f(-\mu^2+i\epsilon)\over 2i}~.
\ee
To get \reef{c0} one needs to take trace of \reef{masscorr}, use Green's equation for $G(x,\mu)$ and  implement the Kramers--Kronig relations to invert the resulting identity. In this Appendix we present details of the computation of \reef{c0} in the case of free massive fields. 

We start with the free Dirac field. The Euclidean action is given by
\be
S=\int d^dx \, \bar\psi\( \slashed\del+m\)\psi~,
\ee
and the corresponding energy-momentum tensor reads
\be
 T_{\mu\nu}={1\over 2} \bar\psi\gamma_{(\al}\overset{\leftrightarrow}{\partial}_{\bt)}\psi-\delta_{\al\bt}(\bar\psi\slashed\del\psi+m\bar\psi\psi).
\ee
Taking the trace of the energy-momentum tensor and using the Dirac equations of motion we obtain,
\be
T=-m\,\bar\psi\psi.
\ee 
Hence, \reef{c0} boils down to a computation of the standard fermionic loop diagram,
\be
 c^{(0)}_F(\mu)=-{2\,\Gamma(d)\over \pi A_d} \, {m^2\over \mu^3} \, \text{Im}\(\text{tr} \int {d^dq\over (2\pi)^d} 
 {\big(-i(\slashed{p}+\slashed{q})+m\big)\big(-i\slashed{q}+m\big)\over \big( (p+q)^2+m^2 \big)\big( q^2+m^2 \big) }\) \Bigg|_{p^2=-\mu^2}~,
 \label{cF2}
\ee 
where ``$\text{tr}$'' denotes the trace over the $2^{[{d\over 2}]}$-dimensional spinor space and we used the Euclidean propagator,
\be
\langle \psi(x) \bar\psi(0)\rangle=(\slashed\del+m)^{-1}=\int{d^dp\over (2\pi)^d} \, {-i\slashed{p}+m\over p^2+m^2} \, e^{ip\cdot x} ~.
\ee

Implementing now the standard Feynman parametrization to evaluate the integral within parenthesis in \reef{cF2} yields,
\be
 c^{(0)}_F(\mu)=2^{[{d\over 2}]+1}(d-1) {\Gamma(d)\over \pi A_d} \, {\Gamma\({2-d\over 2}\)\over (4\pi)^{d/2}}\,{m^2\over \mu^3} \, 
 \text{Im}\( \int_0^1dx \, \sigma^{d-2\over 2} \) \Bigg|_{p^2=-\mu^2}~,
 \label{c0f}
\ee 
where $\sigma=m^2+x(1-x)p^2$.  To evaluate the imaginary part of the remaining integral, we assume that $d=d_0+\epsilon$ with even $d_0$ and expand \reef{c0f} in $\epsilon$. Substituting
\bea
 \Gamma\Big({2-d\over 2}\Big)&=&{2 (-)^{d_0\over 2}\over \Gamma\({d_0\over 2}\)} ~{1\over \epsilon}+\ldots~,
 \non
 \non
 \sigma^{d-2\over 2}&=&\sigma^{d_0-2\over 2}(1+{\epsilon\over 2}\log\sigma+\ldots)~,
 \label{exp}
\eea
one can see that the leading divergent term in \reef{c0f} is always real and therefore does not contribute to $c^{(0)}_F(\mu)$, whereas the next-to-leading term will have a non-trivial imaginary part provided that $\sigma< 0$, \ie if
\bea
 x_-< x < x_+~,
\quad
x_{\pm}={1\over 2} \pm {1\over 2} \sqrt{1-4{m^2\over \mu^2}}~, 
\quad
\mu> 2 m~.
\eea
In particular, 
\be
\text{Im} \, \log\sigma \big|_{p^2=-\mu^2}= - \pi ~,
\ee
and
 \be
  \int_{x_-}^{x_+}dx \(x(1-x)-{m^2\over\mu^2}\)^{d_0-2\over 2} = {\Gamma\({d_0\over 2}\)^2\over \Gamma(d_0)} \(1-4{m^2\over\mu^2}\)^{d_0-1\over 2}~.
 \label{Imint}
 \ee
Substituting these results into \reef{c0f}, we recover \reef{cF}. 

We now repeat the above derivation for the massive minimally coupled scalar field. In this case the energy-momentum tensor is given by 
\begin{equation} 
T_{\mu \nu} = \partial_{\mu} \phi \partial_{\nu} \phi - \frac{1}{2} \delta_{\mu \nu} \Big( (\partial \phi)^2 + m^2\phi^2 \Big)~.
\end{equation}
Hence
\be
 T={2-d\over 2} (\partial \phi)^2 - {d\over 2} m^2\phi^2.
\ee
Using now 
\be
 \langle  \phi(x)\phi(0) \rangle=\int {d^dp\over (2\pi)^d} {e^{ipx}\over p^2+m^2} 
\ee
to evaluate various scalar one loop integrals, yields 
\bea
 \langle  \phi^2 \, \phi^2 \rangle&=&2{\Gamma\({4-d\over 2}\)\over (4\pi)^{d/2}}\int_0^1dx \, \sigma^{d-4\over 2}~,
 \non
 \langle  \big(\del\phi\big)^2 \, \phi^2 \rangle &=&{\Gamma\({2-d\over 2}\)\over (4\pi)^{d/2}}\int_0^1dx \, \sigma^{d-4\over 2}
 \Big(  2(d-1)\sigma - (d-2)m^2  \Big) ~,
 \\
  \langle  \big(\del\phi\big)^2 \, \big(\del\phi\big)^2 \rangle &=&-{\Gamma\({2-d\over 2}\)\over (4\pi)^{d/2}}\int_0^1dx \, \sigma^{d-4\over 2}
 \Big(  4(d+1)\sigma^2-(p^2+4d \, m^2)\sigma+(d-2)m^4 \Big) ~,
 \nonumber
\eea
where as before we used Feynman parametrization to carry out the integrals. Using \reef{exp}-\reef{Imint}, we obtain
\bea
 \text{Im} \langle  \phi^2 \, \phi^2 \rangle\big|_{p^2=-\mu^2}&=& \mu^{d-4}{2\pi \over (4\pi)^{d\over 2}} {\Gamma\(d/2-1\)\over \Gamma(d-2)}
 \(1-4{m^2\over \mu^2}\)^{d-3\over 2} \Theta(\mu-2m)~,
 \non
 \text{Im} \langle  \big(\del\phi\big)^2 \, \phi^2 \rangle \big|_{p^2=-\mu^2}&=& \mu^{d-2}{\pi \over (4\pi)^{d\over 2}} {\Gamma\(d/2-1\)\over \Gamma(d-2)}
 \(1-4{m^2\over \mu^2}\)^{d-3\over 2}\(1-2{m^2\over \mu^2}\) \Theta(\mu-2m)~,
 \non
  \text{Im}  \langle  \big(\del\phi\big)^2 \, \big(\del\phi\big)^2 \rangle \big|_{p^2=-\mu^2} &=& \mu^{d}{\pi \over 2(4\pi)^{d\over 2}} {\Gamma\(d/2-1\)\over \Gamma(d-2)}
 \(1-4{m^2\over \mu^2}\)^{d-3\over 2} \(1-2{m^2\over \mu^2}\)^2\Theta(\mu-2m)~,
 \nonumber
\eea
Combining leads to
\be
c^{(0)}_M(\mu)=\mu^{d-3}{8(d-1)(d+1) \over \Omega_d^2} \(1-4{m^2\over \mu^2}\)^{d-3\over 2}\({d-2\over 4}+ {m^2\over\mu^2}\)^2\Theta(\mu-2m)~.
\ee

\bibliographystyle{utcaps}
\bibliography{AllRef}  
\end{document}